\documentclass[a4paper]{jpconf}
\usepackage{graphicx}
\begin{document}
\newcommand{\bea}   {\begin{eqnarray}}
\newcommand{\eea}   {\end{eqnarray}}
\title{Symmetries of the Schr\"odinger Equation and Algebra/Superalgebra Duality}

\author{Francesco Toppan}

\address{CBPF, Rua Dr. Xavier Sigaud 150, cep 22290-180, Rio de Janeiro (RJ), Brazil}

\ead{toppan@cbpf.br}

\begin{abstract}
Some key features of the symmetries of the Schr\"odinger equation that are common to a much broader class of dynamical systems (some under construction) are illustrated.  I discuss the algebra/superalgebra duality involving first and second-order differential operators. It provides different viewpoints for the spectrum-generating subalgebras. The representation-dependent notion of on-shell symmetry is introduced. The difference in associating the time-derivative symmetry operator with either a root or a Cartan generator of the $sl(2)$ subalgebra is discussed.
In application to one-dimensional Lagrangian superconformal sigma-models it implies superconformal actions which are either supersymmetric or non-supersymmetric.
\end{abstract}

\section{Introduction}
Focusing on simple examples, I illustrate some general features that apply to a vast class of theories including
non-relativistic Schr\"odinger equations in $1+d$ dimensions, the more general invariant  equations associated with ${\ell}$-conformal Galilei algebras, the $D=1$ Lagrangian (super)conformal models (one-dimensional sigma-models), together with several extended supersymmetric versions of these theories.\par
This broad class of dynamical equations  share the properties that are discussed here. For simplicity I review these features in application to the Schr\"odinger equation in $1+1$-dimension with the three choices of the potential (constant, linear and quadratic) which induce the Schr\"odinger algebra [1] as the symmetry algebra of first-order differential operators. On the sigma-model side, I illustrate the simplest $osp(1|2)$-invariant case [2] with one bosonic and one fermionic time-dependent field.\par
The key issues that I am pointing out are the following. We have that $2$ of the $6$ first-order Schr\"odinger operators, the ones generating the Heisenberg-Lie algebra, have a natural  half-integer grading. By taking their anticommutators one can construct $3$ second-order differential operators.  The total number of $9$ operators so constructed define a finite closed structure, which can be either Lie-algebraic (by taking their commutators)
or super-Lie algebraic, in terms of the (anti)-commutators which respect the ${\bf{Z}}_2$ grading. The fact that two compatible structures can be defined for the same set of operators is referred to as ``algebra/superalgebra" duality (it is worth pointing out that the use of higher-order differential operators entering a universal enveloping algebra, together with their relations with higher spin theories, was also advocated in [3]; however, an important feature is that finite Lie or super-Lie algebras are recovered when we include only second-order differential operators of special type). \par
The operator $\Omega$ which defines the Schr\"odinger dynamics, being a second-order differential operator, does not belong to the Schr\"odinger algebra, but to the enlarged algebra (either its Lie-algebraic or its super-Lie algebraic version).   The operator $\Omega$ does not commute with the first-order Schr\"odinger operators. On the other hand, {\em in the given representation}, the {\em on-shell} closure of the algebra is guaranteed once taking into account the $\Omega\Psi=0$ dynamical equation (see Section {\bf 3}).\par
The algebra/superalgebra duality, applied to the spectrum generating subalgebras, is a reformulation of the celebrated result by Wigner in [4]. Essentially, for the harmonic oscillator, the Lie algebra side is based on the Heisenberg subalgebra and the construction of the eigenstates starts with the Fock's vacuum condition.
In the dual super-Lie algebraic point of view, the same conditions are obtained from a highest weight representation of the $osp(1|2)$ spectrum-generating superalgebra.  \par
For the free particle case, the operator $\Omega$ and the time-derivative operator belong to the grading $1$ sector of the algebra (the time-derivative operator is a root of the $sl(2)$ subalgebra). In the harmonic oscillator case $\Omega$ and the time-derivative operator belong to the grading $0$ sector of the algebra
(the time-derivative operator is the Cartan of the $sl(2)$ subalgebra). This is the key difference which implies the continuum spectrum for the free particle and the discrete spectrum of the harmonic oscillator. In [2] a detailed analysis of one-dimensional (super)conformal models based on parabolic $D$-module reps (the time-derivative operator being associated with the positive $sl(2)$ root) versus hyperbolic/trigonometric $D$-module reps (the time-derivative operator being associated with the  $sl(2)$ Cartan generator) was given.
In the hyperbolic/trigonometric case extra potential terms are allowed. On the other hand, when invariance under superconformal algebras are considered, the difference is even more meaningful. In the parabolic case the dynamical system is both superconformal and supersymmetric. In the hyperbolic/trigonometric case the dynamical system is superconformal but not supersymmetric. This point is illustrated in Section {\bf 5}.\par
In the Conclusions I point out how the features here discussed enter more general dynamical systems. These features can be used to both identify and solve the invariant dynamics of this larger class of theories.

\section{Schr\"odinger's equations in $d=1$}
~\par

We consider the differential operator $\Omega$ in $1+1$ dimensions:
\bea
\Omega &=&\partial_t +a \partial_x^2 - a V(x).
\eea
$V(x)$ is a potential term. If $a$ is imaginary, the operator $\Omega$ defines the dynamics of the $1+1$ Schr\"odinger equation, written as
\bea\label{scheq}
\Omega\Psi(x,t)&=& 0.
\eea
If $a$ is real the above equation is the heat equation in the presence of the potential $V(x)$.\par
With standard methods, see [1], we can prove that, for three special cases of the potential, the invariance algebra of the equation (\ref{scheq}), in terms of first-order differential operators, is given by the Schr\"odinger algebra
($l=\frac{1}{2}$ conformal Galilei algebra) in the presence of a central charge. \par
Indeed, the invariant condition $\Omega\delta\Psi(x,t)=0$, \\for $
\delta \Psi(x,t)= f(x,t)\Psi_t(x,t)+g(x,t)\Psi_x(x,t)+h(x,t)\Psi(x,t)$, \\
leads to the set of equations
\bea
f_x&=&0,\nonumber\\
f_t-2g_x+af_{xx}&=&0,\nonumber\\
g_t+a\left(2aVf_x+2h_x+g_{xx}\right)&=&0,\nonumber\\
agV_x+h_t+2a^2V_xf_x+aV(f_t+af_{xx})+ah_{xx}&=&0.
\eea
The three special cases correspond to the solutions with maximal number of generators:\\
{\em i}) the constant potential $V(x)=0$ (free particle case),\\
{\em ii}) the linear potential $V(x) = \omega x$ and\\
{\em iii}) the quadratic potential $V(x)=\nu^2x^2$ (harmonic oscillator case).\par
In all the above cases, without loss of generality, the potential can be shifted by a constant $u$, $V(x)\rightarrow V(x)+u$, via a similarity transformation $\Omega\rightarrow e^{aut}\Omega e^{-aut}$. \par
In the quadratic case a linear term in the potential can always be eliminated by shifting the space coordinate, so that $x \rightarrow x+b$. \par
For all the other potentials, the invariance algebra of eq. (\ref{scheq}) is smaller than the Schr\"odinger algebra.\par
The quadratic and constant realizations can be mutually recovered, see [5], from similarity transformations coupled with change of space and time coordinates. The existence of this set of transformations, however, is not essential for the following discussion.\par
A compatible assignment of the dimensions is
\bea
& [t]=-1,\quad [x]=-\frac{1}{2},\quad [\omega]=\frac{3}{2},\quad [\nu]=1, \quad ([\Omega]=1).&
\eea
The Schr\"odinger algebra is given by the $6$ generators $z_{\pm1}, z_0, w_{\pm}, {c}$. Their dimensions are
\bea\label{grading}
& [z_{\pm 1}]=\pm 1, \quad [w_\pm]=\pm\frac{1}{2},\quad [z_0]=[{c}]=0.
\eea
The generators $z_{\pm 1}, z_0$ close an $sl(2)$ subalgebra with $z_0$ as the Cartan element. The generator ${ c}$ is the central charge.\par
The three explicit $D$-module reps are given by\par
{\em i}) $V(x)=0$, constant potential case,
\bea
z_{+1}&=& \partial_t,\nonumber\\
z_0&=&t\partial_t+\frac{1}{2}x\partial_x +\frac{1}{4},\nonumber\\
z_{-1}&=& t^2\partial_t+tx\partial_x-\frac{x^2}{4a}+\frac{1}{2}t,\nonumber\\
w_+&=& \partial_x,\nonumber\\
w_-&=& t\partial_x-\frac{x}{2a},\nonumber\\
{ c}&=&1;
\eea
{\em ii}) $V(x)=\omega x$, linear potential case,
\bea
z_{+1}&=& t^2\partial_t+(a^2\omega t^3+tx)\partial_x+(\frac{t}{2}-\frac{1}{4}a^3\omega^2t^4-\frac{3}{2}a\omega t^2x-\frac{x^2}{4a}),\nonumber\\
z_0&=&-t\partial_t-(\frac{3}{2}a^2\omega t^2+\frac{x}{2})\partial_x+(
\frac{1}{2}a^3\omega^3t^3+\frac{3}{2}a\omega t x-\frac{1}{4}),\nonumber\\
z_{-1}&=& \partial_t+2a^2\omega t\partial_x-a^3\omega^2t^2-a\omega x,\nonumber\\
w_+&=& -t\partial_x+\frac{x}{2a}+\frac{1}{2}a\omega t^2,\nonumber\\
w_-&=& \partial_x-a\omega t,\nonumber\\
{c}&=&1;
\eea
{\em iii}) $V(x)=\nu^2x^2$, quadratic potential case
\bea
z_{+1}&=& e^{4a \nu t}\left(\partial_t +2a\nu x\partial_x + a\nu -2 a\nu^2x^2\right),\nonumber\\
z_0&=&\partial_t,\nonumber\\
z_{-1}&=&e^{-4a \nu t}\left(\partial_t -2a\nu x\partial_x - a\nu -2 a\nu^2x^2\right),\nonumber\\
w_+&=& e^{2a\nu t}\left(\partial_x-\nu x\right),\nonumber\\
w_-&=& e^{-2a\nu t}\left(\partial_x+\nu x\right),\nonumber\\
{ c}&=&1.
\eea
In the quadratic case the non-vanishing commutation relations are given by 
\bea\label{schalg}
\relax [ z_1,z_{-1}] &=& -8 a\nu z_0,\nonumber\\
\relax [ z_{0 }, z_{\pm 1}] &=& \pm 4 a\nu z_{\pm 1},\nonumber\\
\relax [z_{\pm 1}, w_\mp] &=& \mp 4 a \nu w_\pm,\nonumber\\
\relax [z_0, w_\pm] &=& \pm 2 a \nu w_\pm,\nonumber\\
\relax [w_+,w_-] &=& 2\nu { c}.
\eea
In the constant and linear cases the commutation relations are obtained from the above formulas with the substitution $\nu =-\frac{1}{4a}$. \par
The above equations give the structure constants of the one-dimensional, centrally extended, Schr\"odinger algebra.\par
The generator $z_0$ defines the grading corresponding to the (\ref{grading}) dimensions.\par
One should note that the Hamiltonian (i.e., the time-derivative operator), corresponds to a grading $1$ generator (a root generator of the $sl(2)$ subalgebra) in the free particle case and to a grading $0$ generator
(the Cartan generator of the $sl(2)$ subalgebra) for the harmonic oscillator case.\par
The generators $w_\pm$ have half-integer grading with respect to the grading defined by $z_0$.

\section{The algebra/superalgebra symmetry with higher differential operators}

The second-order differential operators $w_{1}, w_0, w_{-1}$, obtained by taking the anticommutators of
$w_\pm$, can be constructed:
\bea
w_{+1} &=& \{w_+,w_+\},\nonumber\\
w_0&=& \{w_+,w_-\},\nonumber\\
w_{-1} &=& \{w_-,w_-\}.
\eea
Their explicit form, in the three respective cases above, is given by\par
{\em i}) the constant case,
\bea
w_{+1}&=& 2{\partial_x}^2,\nonumber\\
w_0 &=& 2t{\partial_x}^2-\frac{x}{a}\partial_x -\frac{1}{2a},\nonumber\\
w_{-1} &=& 2t^2{\partial_x}^2-\frac{2tx}{a}\partial_x +\frac{x^2}{2a^2}-\frac{t}{a};
\eea
{\em ii}) the linear case,
\bea
w_1&=&\frac{1}{2a^2}\left( 4a^2t^2{\partial_x}^2-4at(a^2t^2\omega+x)\partial_x+
(a^4t^4\omega^2-2at+2a^2t^2\omega x+x^2)\right),\nonumber\\
w_0&=& \frac{1}{2a}\left(4at{\partial_x}^2-(6a^2t^2\omega+2x)\partial_x+(2a^3t^3\omega^2+2at\omega
x-1)\right),\nonumber\\
w_{-1} &=& 2{\partial_x}^2-4at\omega\partial_x +2a^2t^2\omega^2;
\eea
{\em iii}) the quadratic case,
\bea
w_{+1}&=&e^{4at\nu}\left(2{\partial_x}^2-4\nu x\partial_x-2\nu +2\nu^2x^2\right),\nonumber\\
w_0&=& 2{\partial_x}^2-2\nu^2x^2,\nonumber\\
w_{-1}&=& e^{-4at\nu}\left(2{\partial_x}^2+4\nu x\partial_x+2\nu+2\nu^2x^2\right).
\eea
In all these cases we have two consistent closed structures which can be defined on the same set of differential operators, namely\\
{\em 1}) the non-simple Lie algebra $eSch$ (the enlarged Schr\"odinger algebra), presented by the  $9$ generators
$\{z_{\pm 1}, z_{0}, w_{\pm}, w_{\pm 1}, w_0, {c}\}$ and\\
{\em 2}) the Lie superalgebra $sSch$ (the enlarged Schr\"odinger superalgebra) $sSch= S_0\oplus S_{1}$, with $7$ even generators
($ z_{\pm1}, z_0, w_{\pm 1}, w_0, { c}\in S_0$) and $2$ odd generators ($w_\pm \in S_1$).\par

In all three cases (in the quadratic case for $\nu =-\frac{1}{4a}$), the extra non-vanishing structure constants besides (\ref{schalg}) are given,
for the $eSch$ algebra, by
\bea
\relax [z_0, w_{\pm 1}]&=&\mp w_{\pm1},\nonumber\\
\relax [z_{\pm 1}, w_0]&=& \pm w_{\pm 1},\nonumber\\
\relax [z_{\pm 1}, w_{\mp 1} ] &=& \pm 2 w_0,\nonumber\\
\relax [w_\pm, w_0]&=& \mp \frac{1}{a}w_{\pm},\nonumber\\
\relax [w_\pm, w_{\mp 1}] &=& \mp \frac{2}{a}w_\mp,\nonumber\\
\relax [ w_0, w_{\pm 1}] &=& \pm \frac{2}{a} w_{\pm 1},\nonumber\\
\relax [ w_1,w_{-1}] &=& -\frac{4}{a}w_0.
\eea

For the $sSch$ superalgebra we have the anti-commutators
\bea
&\{w_+,w_+\}= w_{+1},\quad \{w_+,w_-\}= w_0,\quad \{w_-,w_-\}= w_{-1}&
\eea
(one should note that the Heisenberg-Lie algebra $[w_+,w_-]=2\nu { c}$ is not a subalgebra of the $sSch$ superalgebra).\par
In all three cases (constant, linear and quadratic), the second-order differential operator $\Omega$ is a generator belonging to the enlarged Schr\"odinger algebra (either $eSch$ or $sSch$). We have the following identifications:\\
{\em i}) constant case,
\bea
\Omega &=& z_{+1}+\frac{1}{2}a w_{+1},
\eea
{\em ii}) linear case,
\bea
\Omega &=& z_{-1}+\frac{a}{2}w_{-1},
\eea
{\em iii}) quadratic case,
\bea
\Omega &=& z_0+\frac{a}{2}w_0.
\eea
In each case, either the $eSch$ algebra or the $sSch$ superalgebra, is the {\em on-shell} symmetry algebra
of the evolution equation determined by $\Omega$. We have indeed that\\
{\em i}) in the constant case, all commutators involving $\Omega$ are vanishing, apart from
\bea
\relax [z_0,\Omega] &=& - z_{+1}-\frac{a}{2} w_{+1},\nonumber\\
\relax  [z_{-1}, \Omega ] &=& -2 z_0-aw_0.
\eea
 In the given representation, on the other hand, the above commutators are identified with the representation-dependent formulas
\bea
\relax [z_0,\Omega ] &=& -\Omega,\nonumber\\
\relax [z_{-1},\Omega] &=& -2t\Omega.
\eea
{\em ii}) in the linear case all commutators involving $\Omega$ are vanishing, apart from
\bea
\relax [z_{+1},\Omega ]&=& 2z_0+aw_0,
\nonumber\\
\relax [z_0,\Omega ]&=& z_{-1}+\frac{a}{2}w_{-1}.
\eea
 In the given linear representation, on the other hand, we have
\bea
\relax  [z_{+1},\Omega] &=& 2t \Omega,\nonumber\\
\relax [z_0,\Omega ] &=&\Omega .
\eea
{\em iii}) In the quadratic case all commutators involving $\Omega$ are vanishing, apart from
\bea
\relax [z_{+1},\Omega ] &=& z_{+1}+\frac{1}{2} w_{+1},\nonumber\\
\relax [z_{-1},\Omega ] &=& - z_{-1}-\frac{1}{2}a w_{-1}.
\eea 
In the given quadratic representation, on the other hand, we have
\bea
\relax [z_{+1},\Omega ]&=& e^{-t}\Omega,\nonumber\\
\relax [z_{-1},\Omega ] &=& -e^{t}\Omega.
\eea
Comment: the fact that we obtain the representation-dependent commutators
\bea
\relax [g,\Omega ] &=& f_g\cdot \Omega,
\eea
for any generator $g$ of either the $eSch$ or the $sSch$ algebra, with $f_g$ a given function, tells us that
$eSch$ or $sSch$ is the on-shell symmetry (super)algebra for the $\Omega\Psi(t,x)=0$ equation.
 
\section{Duality for spectrum generating algebras/superalgebras}

The famous Wigner's analysis in [4], which allows in particular to solve the harmonic oscillator without
using the canonical commutation relations, can be understood from the results discussed in the previous Section. In particular from the notion of algebra/superalgebra duality for the first and second order differential operators closing the on-shell symmetry (super)algebra of the one-dimensional oscillator.\par
We recall that the given set of $9$ differential operators close, on the Lie algebra side, the ${eSch}$ enlarged Schr\"odinger algebra, while on the super-Lie algebra side they induce the $sSch$ superalgebra.\par
On the Lie algebra side the spectrum generating algebra allowing to reconstruct the eigenfunctions and eigenvalues of the harmonic oscillator is the Heisenberg-Lie algebra generated by 
$w_\pm = e^{\pm 2 a \nu t}(\partial_x \pm \nu x)$.\par
The (unnormalized) vacuum solution $\Psi_{vac}(x,t)$ of the $\Omega\Psi(x,t)=\Psi_t+a\Psi_{xx}-a\nu^2x^2\Psi=0$ equation, on the Lie algebra side, satisfies the Fock's vacuum condition 
\bea\label{fock}
w_+\Psi_{vac}(x,t)&=&0,
\eea
 together with the equations 
\bea\label{root}
z_0\Psi_{vac}(x,t)&=& -a\nu \Psi_{vac}(x,t), \nonumber\\
w_0\Psi_{vac}(x,t)&=& a\nu \Psi_{vac}(x,t).
\eea
The explicit solution is given by $\Psi_{vac}(x,t) = Ce^{-a\nu t} e^{-\frac{\nu}{2}x^2}$.\par
The eigenstates $\Psi_n(x,t)$,  corresponding to higher energy eigenvalues of the harmonic oscillator, are constructed through the positions
\bea
 \Psi_n(x,t) &=& (w_-)^n \Psi_{vac}(x,t).
\eea
By construction, they satisfy the $\Omega\Psi_n(x,t)=0$ equation.\par
From the dual, superalgebraic, point of view, we have a spectrum-generating superalgebra given by the simple Lie superalgebra
$osp(1|2)\subset sSch$. Its generators are $w_0, w_{\pm 1}$ and $w_\pm$.\par
In the superalgebra picture the same conditions to reconstruct eigenstates and eigenvalues of the harmonic oscillator are read differently. The Equation (\ref{fock}) and the second equation in (\ref{root}) define a highest weight representation of $osp(1|2)$, with $\Psi_{vac}(x,t)$ being its highest weight vector.
\section{Other cases: supersymmetric versus non-supersymmetric superconformal mechanics}

The symmetry operator of the Schr\"odinger equation expressed via the time derivative corresponds, in the free particle case, to the Cartan generator of $sl(2)$ and, in the oscillatorial case, to a root generator of the $s(2)$-invariant subalgebra.\par
This feature is also present in other different contexts. In particular, in the case of (super)conformal mechanics
in $0+1$ dimensions, realized in the Lagrangian setting.\par
Based on some results in [6], it was shown in [2] that the $D$-module reps of the (super)conformal algebras admit parabolic as well as hyperbolic/trigonometric realizations.  These transformations define superconformally invariant actions. In the hyperbolic/trigonometric case, extra potentials, not allowed in the parabolic case, are present. This one, on the other hand, is not the only difference concerning the various types of realizations. In the parabolic case, the time-derivative operator (i.e., the
``Hamiltonian") is associated with a positive root of the $sl(2)$-invariant subalgebra, while in the  
trigonometric/hyperbolic case it is associated with the Cartan element.  As a consequence, we obtain different classes of superconformally-invariant models. In the parabolic case, the Hamiltonian, being  associated to a bosonic root, is  the square of the fermionic symmetry operators related to the simple fermionic roots. The resulting theory, besides being superconformal, is also supersymmetric in the ordinary sense of the word ``supersymmetry". A different picture emerges in the hyperbolic/trigonometric case. The Hamiltonian is still a symmetry operator. On the other hand, it cannot be expressed as a square of fermionic symmetry operators. The resulting theory is superconformally-invariant, but not supersymmetric. Alternatively (following [7], which proposed this term in a different context), we can introduce the notion of ``weak supersymmetry" to refer to this feature.\par
Indeed, the ${\cal N}$-extended ordinary supersymmetry requires, for a given ${\cal N}$, that a set of ${\cal N}$ fermionic
symmetry generators $Q_i$ closes the supersymmetry algebra $\{Q_i,Q_j\}=2\delta_{ij}H$, $[H,Q_i]=0$ ($i,j=1,\ldots, {\cal N}$), where $H$ is the time-derivative operator (the ``Hamiltonian").\par
In the hyperbolic/trigonometric cases, ${\cal N}$ fermionic symmetry generators can be found. They are the square roots of a symmetry generator (let's call it $Z$), which does not
coincide with the Hamiltonian $H$. As a matter of fact, in the hyperbolic/trigonometric cases, two independent symmetry subalgebras $\{Q_i^\pm,Q_j^\pm\}=2\delta_{ij}Z^\pm$, $[Z^\pm,Q_i^\pm]=0$ (with
$Z^+\neq H$ and $Z^-\neq H$) are encountered. In the parabolic cases two independent symmetry subalgebras are also encountered and one of them can be identified with the
ordinary supersymmetry ($Z^-= H$, $Z^+\neq H$).\par
In the hyperbolic/trigonometric cases the Hamiltonian $H$ continues to be a symmetry operator. It belongs, however, to the $0$-grading sector of the superconformal algebra and  is not
the square of any fermionic symmetry operator (contrary to the operators $Z^\pm$, which belong to the $\pm 1$ grading sectors, respectively).\par
These points are conveniently illustrated with the simplest example, the ${\cal N}=1$ theory based on the $(1,1)$ supermultiplet (a single bosonic field $\varphi$ and a single fermionic field $\psi$) admitting constant kinetic term and $osp(1|2)$ invariance. The action can be written as
\bea\label{hypern1}
{\cal S} &=& \int dt ({\dot\varphi}^2 -\psi{\dot \psi}+\epsilon\varphi^2).
\eea
The potential term is absent ($\epsilon=0$) in the parabolic realization of the superconformal invariance. It is present in the hyperbolic ($\epsilon =1$) and in the trigonometric  ($\epsilon= -1$) realizations. In the hyperbolic case
the five invariant operators (closing the $osp(1|2)$ algebra) are given by
\bea
Q^{\pm} \varphi = e^{\pm t} \psi, && Q^\pm \psi = e^{\pm t} ({\dot\varphi}\mp \varphi),\nonumber\\
Z^\pm \varphi =e^{\pm 2 t} ({\dot\varphi}\mp \varphi), && Z^\pm \psi = e^{\pm 2t}{\dot \psi},\nonumber\\
H \varphi ={\dot\varphi}, && H\psi = {\dot \psi}.
\eea
One should note that $Z^\pm=({Q^\pm})^2$.\par
No change  of time variable $t\mapsto \tau (t)$ allows to represent either $Z^+$ or $Z^-$ as a time-derivative operator with respect to the new time $\tau$.

\section{Conclusions} 

The several key features discussed in this paper can be extended to investigate the dynamics of more complicated systems. The algebra/superalgebra duality involving a finite number of first-order and second-order differential
operators can be constructed not only only for Schr\"odinger equations in $1+d$-dimensions, but also from ${\ell}$-extended conformal Galilei algebras (a discussion of these first-order differential operator algebras can be found in [8], [9] and [10]), with half-integer ${\ell}$ (${\ell}=\frac{1}{2}$ corresponds to the Schr\"odinger algebra). The generators associated with the half-integer grading can be naturally identified with 
the odd generators in the superalgebra picture. The representation-dependent notion of on-shell symmetry (as discussed in Section {\bf 3})
is applicable to construct the invariant dynamics associated with these conformal Galilei algebras. Invariant operators exist both at grading $1$ (the generalization of the Schr\"odinger free case) and grading $0$
(the generalization of the Schr\"odinger oscillatorial case). Unlike the Schr\"odinger case, where the invariant equation is given and the symmetry operators are derived with standard techniques, an inverse problem is defined. The algebra is now given; it is the invariant operator induced by the given representation that has to be computed.
As a result we can identify new solvable differential equations.
A joint paper with N. Aizawa and Z. Kuznetsova concerning this construction is currently under finalization.\par
Another feature which deserves to be noticed is that, if starting from a supersymmetric system, the 
construction leading to the algebra/superalgebra duality is replaced by a construction leading to superalgebra/ $({\bf Z}_2\times{\bf Z}_2)$-graded algebra duality, where the notion of ${\bf Z}_2\times{\bf Z}_2$-graded superalgebra can be found, e.g., in [11]. Another paper about this construction is currently under finalization.\par
On the side of supersymmetric and non-supersymmetric superconformal one-dimensional sigma models, at present the quantization of these classical systems is under investigation.

\par~\par
\subsection{Acknowledgments}

This paper received support from CNPq through  the PQ grant 306333/2013-9 .\\
I profited of useful discussions with N. Aizawa, N. L. Holanda, Z. Kuznetsova and M. Valenzuela.

\end{document}